\documentclass[epj,nopacs]{svjour}
\usepackage{graphics}
\begin{document}
\title{Open and Hidden Charm Production in d$A$ Collisions at RHIC and LHC}
\author{R. Vogt}
\institute{Lawrence Berkeley National Laboratory, Berkeley, CA 94720, USA 
\and 
Physics Department, University of California, Davis, CA 95616, USA
}
\date{Received: date / Revised version: date}
%
\abstract{We discuss aspects of open and hidden charm production in 
deuterium-nucleus
collisions at RHIC and LHC energies.  
We describe calculations of the total $c \overline c$ cross section and the
charm quark transverse momentum distributions. We next explain how shadowing
and moderate nuclear absorption can explain the PHENIX $J/\psi$ dAu/$pp$
ratios and predict the combined effect of shadowing and absorption in 6.2 TeV
d+Pb collisions.
} 
\maketitle
\section{Introduction}
\label{intro}

Heavy flavors were one of the first true hard probes of heavy ion collisions
through the $J/\psi$ measurements of NA38 and its successors 
\cite{Klubergproc}.  These same measurements also helped show the importance
of baseline measurements in $pp$ and $pA$ interactions to separate `normal'
from `anomalous' behavior.  Now, after initial Au+Au runs at RHIC, the first
baseline $pp$ and d+Au data have been reported for heavy flavors, both open
charm and $J/\psi$.

Open charm decay leptons were separated for the first time in a heavy ion
environment by PHENIX in Au+Au collisions at $\sqrt{S_{NN}}=130$ GeV
\cite{PHENIX130}.  Now STAR has results 
both in the lepton channel \cite{startalk}
and in the spectra of reconstructed $D$ mesons \cite{Marcoproc}.  These 
measurements will help more fully understand open charm production in $AA$
collisions where energy loss \cite{magdaproc}, flow \cite{rappproc} and 
regeneration of $J/\psi$ by $c \overline c$ \cite{thewsproc} may be important.
The $J/\psi$, on the other hand, while well measured at the SPS, awaits data 
from the 2004 Au+Au run to provide a higher statistics measurement of
$J/\psi$ production at 200 GeV.  The d+Au $J/\psi$ data, over a wide rapidity
range, provide interesting hints of the behavior of the nuclear gluon 
distribution.  We discuss our d$A$ calculations at RHIC and provide predictions
for $J/\psi$ production in d+Pb collisions at the LHC.

\section{Open Charm}
\label{open}

Open charm measurements date back to the late 1970s when $D$ 
and $\overline D$ mesons were first detected, completing the
picture of the fourth quark begun when the $J/\psi$ was detected in $p$Be
and $e^+ e^-$ interactions.  The charm quark was postulated to have a mass
between 1.2 and 1.8 GeV, calculable in perturbative quantum 
chromodynamics (pQCD).  Because of its relatively large mass, it is 
possible to calculate a total $c \overline c$ cross section, not
the case for lighter flavors.  Charm hadrons are usually
detected two ways.  The reconstruction of charged hadron decays 
such as $D^0 \rightarrow K^- \pi^+$ (3.8\%) and $D^+ \rightarrow K^- \pi^+ 
\pi^+$ (9.1\%) gives the full momentum of the initial $D$ meson, yielding
the best direct measurement.  Charm can also be detected indirectly via
semi-leptonic decays such as $D \rightarrow K l \nu_l$ although
the momentum of the parent $D$ meson remains unknown.  Early measurements
of prompt leptons in beam dump experiments assumed that the density of the dump
was high enough to absorb semi-leptonic decays of non-charm hadrons,
leaving only the charm component.  At modern colliders, it is not possible to
use beam dumps to measure charm from leptons but, at sufficiently
high $p_T$, electrons from charm emerge from hadronic cocktails 
\cite{startalk,phenixtalk}.

Extracting the total charm cross section is a non-trivial task.  To go from
a finite number of measured $D$ mesons in a particular decay channel
to the total $c \overline c$ cross section one must: divide by the branching 
ratio; correct for the luminosity, $\sigma_D = N_D/ 
{\cal L}t$; extrapolate to full phase space from the finite detector 
acceptance; divide by two to get the pair cross section from the single $D$s;
and multiply by a correction factor \cite{Mangano} to account for the 
unmeasured charm hadrons, primarily $D_s$ and $\Lambda_c$.  
There are assumptions all along the way.  The most
important is the extrapolation to full phase space.  Before QCD calculations
were available, the data were extrapolated assuming
a power law for the $x_F$ distribution, related to the longitudinal
momentum of the charm hadron by $x_F = p_z/(\sqrt{S}/2) = 2 m_T 
\sinh y/\sqrt{S}$.  The canonical parameterization for extrapolation over all
$x_F$ is $(1 - x_F)^c$ where $c$
was either fit to data over a finite $x_F$ range or simply assumed.  These
parameterizations could lead to large overestimates of the total cross section
if $0<c<2$ was assumed, especially when data were taken only near
$x_F = 0$.  Lepton measurements, particularly using beam dumps, 
resulted in more conservative cross sections
but were typically at more forward $x_F$.

\subsection{Total cross section}

There has been a great deal of improvement over the last 10-15 years.  
Next-to-leading order (NLO) calculations are used in the phase space 
extrapolation, resulting in considerably less ambiguity in the shape of the
$x_F$ distributions, $d\sigma/dx_F$.  The transverse momentum distributions
are more difficult, as we will discuss later.  To calculate the total
cross section to NLO, scaling functions \cite{NDE} proportional to logs of 
$\mu^2/m^2$ are useful where $\mu$ is the scale of the hard process.  
The hadronic cross section in $pp$ collisions can
be written as
\begin{eqnarray}
\sigma_{pp}(S,m^2) & = & \nonumber \\ 
&  & \mbox{} \!\!\!\!\!\!\!\!\!\!\!\!\!\!\!\!\!\!\!\!\!\!\!\!\!\!\!\!\!
\!\!\!\!\!\!\!\!\!\!\!\!\!\!\!\!\!\!\! 
\sum_{i,j}
\int dx_1 \, dx_2 \, 
f_i^p (x_1,\mu_F^2) \,
f_j^p(x_2,\mu_F^2) \, \widehat{\sigma}_{ij}(s,m^2,\mu_F^2,\mu_R^2)
\label{sigpp}
\end{eqnarray}
where the sum over $i,j$ runs over $q$, $\overline q$ and $g$ while
$x_1$ and $x_2$ are the fractional momenta carried by the colliding
partons and $f_i^p$ are the proton parton densities.
The partonic cross sections are
\begin{eqnarray}
\widehat{\sigma}_{ij}(s,m,\mu_F^2,\mu_R^2) & = & 
\frac{\alpha_s^2(\mu_R^2)}{m^2}
\left\{ f^{(0,0)}_{ij}(\rho) \right. \nonumber \\
 & & \!\!\!\!\!\!\!\!\!\!\!\!\!\!\!\!\!\!\!\!\!\!\!\!\!\!\!\!\!
\!\!\!\!\!\!\!\!\!\!\!\!\!\!\!\!\!\!\!\!\!\!\! 
\left. + 4\pi \alpha_s(\mu_R^2) \left[f^{(1,0)}_{ij}(\rho) + 
f^{(1,1)}_{ij}(\rho)\ln\bigg(\frac{\mu_F^2}{m^2} \bigg) \right] \!
+ \! {\cal O}(\alpha_s^2) \right\}
\,\, .
\label{sigpart}
\end{eqnarray}
with $s$ the squared partonic center of mass energy, $\rho = 4m^2/s$ and 
$f_{ij}^{(k,l)}$ are the scaling functions given to NLO in Ref.~\cite{NDE}.
It is most consistent to assume that the factorization scale, $\mu_F$, and
the renormalization scale, $\mu_R$, are equal, $\mu = \mu_F = \mu_R$.
There is no dependence on the kinematic variables.
Some NNLO calculations are available near threshold, $s = x_1 x_2 S \sim 1.3 \,
(4m^2)$, applicable only for $\sqrt{S} \leq 20-25$ GeV 
\cite{KLMVcc,KVcc}.
The NLO corrections to the leading order (LO) cross sections are relatively
large, $K^{(1)} = \sigma_{\rm NLO}/\sigma_{\rm LO} \sim 2-3$, depending on
$\mu$, $m$ and the parton densities \cite{RVkfac}.  

The NNLO corrections are
about as large to next-to-next-to-leading logarithm \cite{KLMVcc} but decrease
to less than $K^{(1)}$ when subleading logs are included \cite{KVcc}.  
This $K$ factor
is large because, in the range $1.2<m<1.8$ GeV, $m <\mu <2m$ with a 5-flavor
QCD scale, $\Lambda_5$, of $0.153$
GeV for the GRV98 HO and $0.22$ GeV for the MRST parton densities,
giving $0.21 < \alpha_s(c) < 0.4$.

\begin{figure*}
\resizebox{0.95\textwidth}{!}{%
  \includegraphics{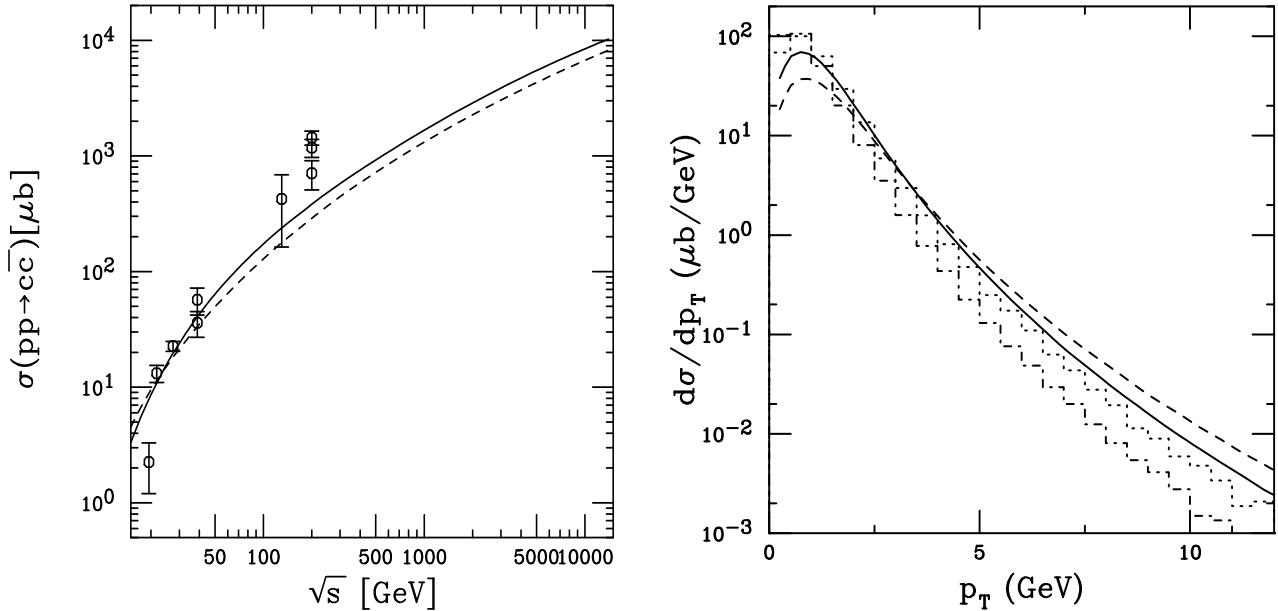}
}
\caption[]{The NLO total $c \overline c$ cross sections as a function of
$\sqrt{S}$ (left-hand side)
and charm quark $p_T$ distribution at $\sqrt{S} = 200$ GeV
in the range $|y| \leq 1$ (right-hand side)
in $pp$ interactions.
The curves are MRST HO (solid) with $m = 1.2$ GeV and $\mu^2 =
4m^2$ and GRV98 HO (dashed) with $m = 1.3$ GeV and $\mu^2 = m^2$.
The two histograms include fragmentation with the Peterson function using
$\epsilon_c = 0.06$ (dot-dashed) and 0.006 (dotted).
}
\label{totpluspt}
\end{figure*}

Instead of presenting a wide range of possible cross sections and emphasizing 
the uncertainties, the approach taken in Ref.~\cite{HPC} has been to 
``fit'' $m$ and $\mu$ for a particular parton density and
extrapolate to higher energies.  The results are compared to some 
of the total cross
section data \cite{Mangano} on the left-hand side of Fig.~\ref{totpluspt}.
The data tend to favor lower values of $m$, $1.2-1.3$ GeV.  The two curves
cross each other because the MRST calculation with $\mu = 2m$ increases faster
at large $\sqrt{S}$ and smaller $x$ due to the stronger QCD evolution of the 
parton densities at the higher scale.  
Although the fixed target results are in good agreement with the
calculations, the PHENIX point \cite{PHENIX130} at 130 GeV, from Au+Au electron
measurements, and the STAR point \cite{STAR}, from a combination of
electron and reconstructed $D$ measurements, are generally above the
calculations.  The STAR point is about
a factor of four over the calculated cross section.  The higher energy
$p \overline p$ data from UA1 \cite{UA1} and CDF \cite{CDF} (not shown, the CDF
data are, in any case, only at high $p_T$, not allowing a total cross section 
measurement) are in better
agreement with the calculations.  

\subsection{Transverse momentum distributions}

Now we turn to the transverse momentum, $p_T$, distributions.  In this case,
the quark mass is no longer the only scale and $p_T$-dependent logs
also appear.  Thus, to interpolate between
the high $p_T$ scale of $p_T$ and the low $p_T$ scale of $m$, a scale
proportional to $m_T$, the transverse mass, is the natural choice.
The charm quark 
$p_T$ distributions are not strongly dependent on quark mass for $p_T \geq 3$
GeV, as may be expected, where the difference in rate
is $\approx 20$\% between $m = 1.2$ and 1.8 GeV.  The
difference in the total cross sections is almost all at $p_T \leq 3$
GeV.  Changing the scale changes the slope of the $p_T$ distributions.  
The distributions are harder for $\mu = m$ than $\mu = 2m$.
The average $p_T$, $\langle p_T \rangle$, increases with $m$ and is larger for
$\mu = m$.  

It is not enough to equate the charm quark and the $D$ meson since 
hadronization, $c \rightarrow D$, may involve some momentum loss by the
quark. If factorization holds
in the final state (universal fragmentation functions) as well as in the
initial state (universal parton distributions) then the fragmentation functions
extracted in $e^+ e^-$ should also be applicable to $pp$ and d$A$.  
However, this assumption typically does not work well for charm \cite{VBH2}. 
The Peterson fragmentation function \cite{Petefun}, 
typically used in hadroproduction codes,
is parameterized as
\begin{eqnarray}
D_{c/H}(z) = N \frac{1}{z} \bigg(1 - \frac{1}{z} - \frac{\epsilon_c}{1-z}
\bigg)^{-2} \, \, 
\label{Pete}
\end{eqnarray}
where $\epsilon_c = 0.06$ was fit to pre-LEP $e^+e^-$ data \cite{Chrin},
reducing the charm hadron momentum by 30\% relative to the charm quark.  
Current measurements by Belle give a slightly smaller value of $\epsilon_c =
0.052$ and also suggest that this functional form gives the worst fit to the 
data of all the fragmentation functions compared to the data \cite{belle}.
Fragmentation functions calculated in Mellin moment, $n$, space
instead of $z$ space, such as the Peterson function, tend to predict lower
momentum loss by the heavy quark, reducing $\epsilon_c$ by up to an order of
magnitude in the $z$-space representation \cite{stefanoproc}.
(In low $\sqrt{S}$ collisions, the momentum reduction due to fragmentation
can be compensated by intrinsic transverse momentum, $k_T$, 
broadening.  However,
such broadening cannot compensate the $x_F$ distributions,
only marginally affected by $k_T$ smearing.  We have previously shown that the
$D$ meson $x_F$ distributions are consistent with no momentum loss
during charm quark hadronization \cite{VBH2}.)  The exclusive NLO 
$Q \overline Q$ code of Ref.~\cite{MNR} includes Peterson 
fragmentation and intrinsic transverse momentum, $k_T$, broadening by adding 
the $k_T$ kick in the final, rather than the initial state.  

The effects of fragmentation and intrinsic $k_T$ broadening of
$\langle k_T^2 \rangle = 1$ GeV$^2$
compensate each other at $\sqrt{S} = 20$ GeV to give a $D$
meson $p_T$ distribution very similar to that of the charm quark.  
This outcome is desirable because the $D$ $p_T$ and $x_F$ distributions are
similar to those of the charm quark at fixed-target energies \cite{VBH2}.
However, at RHIC energies,
due to the higher $\langle p_T \rangle$ at larger $\sqrt{S}$, the effect
of broadening is relatively small and cannot compensate for the momentum loss
induced by fragmentation.  Interestingly enough, the STAR $D$ and $D^*$
$p_T$ distribution also agrees rather well with the NLO charm
quark distribution, as shown in Calderon's talk \cite{Marcoproc}.  On
the right-hand side of Fig.~\ref{totpluspt}, we show the $p_T$
distributions at $\sqrt{S} = 200$ GeV for the two sets of parameters in the
total cross section curves on the left-hand side.  The differences in the 
slopes are due to the different scales while the normalization difference is
due to the choice of charm mass and the parton densities --- the MRST densities
generally give a larger cross section due to their larger $\alpha_s$.  However,
the curves need to be scaled up by a factor of four to agree with the STAR
normalization \cite{Marcoproc}, as may be expected from the 
total cross section results.  

To illustrate the effect of Peterson fragmentation, 
the $p_T$ distribution with $\epsilon_c
= 0.06$ in Eq.~(\ref{Pete}), is shown by the dot-dashed histogram 
in Fig.~\ref{totpluspt}.  The MRST
parton densities are employed, along with $m = 1.2$ GeV, $\mu = 2m$, as in 
the solid curve.  At $p_T \sim 10$ GeV, there is about a factor of 5 between 
the solid curve and the dot-dashed histogram while there is a slight increase
for $p_T \leq 2$ GeV.  If $\epsilon_c$ is decreased by a factor of 10, making
it more consistent with the Mellin space result \cite{stefanopriv}, the
resulting dotted histogram is rather similar to the charm quark
distribution in the solid curve.  A similarly hard fragmentation function is
used in the FONLL formalism, discussed in Frixione's talk \cite{stefanoproc},
which matches fixed-order NLO terms to next-to-leading log, large $p_T$ 
resummation to produce an improved result for $p_T \gg m$ \cite{FONLLinprog}.

The shape of the
charm quark $p_T$ distribution at $\sqrt{S} = 1.96$ TeV
also agrees quite well with the CDF data from the
Tevatron \cite{CDF}.  Given the large discrepancy between the pQCD result and
the STAR cross section, it might be surprising that the normalization is also 
in good agreement with the sum of the charged
and neutral $D$ data scaled to include $D_s$ and $\Lambda_c$ production.
No total cross section is available because only charm
hadrons with $p_T > 5$ GeV have been measured so far.  

Factorization breaking for charm fragmentation has been suggested
from studies of the $x_F$ distributions of {\it e.g.} $D^+$ and $D^-$ 
production,
particularly in $\pi^- A$ interactions where the $D^-$ is leading relative
to the $D^+$ since the $D^-$ shares a valence quark with the $\pi^-$ while the 
$D^+$ does not.
Several mechanisms such as intrinsic charm \cite{VBH2} and string drag have
been proposed, both of which involve charm quark coalescence with spectators.
Such comoving partons are naturally produced in a hard scattering.  Although
it is not intuitive to expect coalescence to work at high $p_T$, it seems
to do so for charm.  See the talks by Hwa \cite{hwaproc} and Rapp 
\cite{rappproc} for discussions of
coalescence models for light hadrons and charm.

\section{Hidden Charm}

We now turn to $J/\psi$ production in d$A$ interactions at RHIC and the LHC.  
It is essential that the $A$ dependence be
understood in cold nuclear matter to set a proper baseline for quarkonium
suppression in $AA$ collisions.  The NA50 collaboration 
\cite{Klubergproc,NA50talks}
has studied the $J/\psi$ $A$ dependence and attributed its behavior to $J/\psi$
break up by nucleons in the final state, referred to as nuclear absorption.
However, it is also known that the parton distributions are modified in
the nucleus relative to free protons.  This modification, referred to here as
shadowing, is increasingly important at higher energies, as we have recently
emphasized \cite{psidaprl}.  In this section, we discuss
the interplay of shadowing and absorption in d+Au collisions at
RHIC and in d+Pb collisions at the LHC.
Previously, we calculated the effect of shadowing alone on the $J/\psi$
d$A$/$pp$ ratio as a function of rapidity and impact parameter \cite{psidaprl}.
The large $c \overline c$ total cross section also has implications for the
$J/\psi$ yield if $J/\psi$'s arise from $c \overline c$ recombination in
a QGP \cite{thewsproc}.  Such a total cross section would 
suggest significant secondary
$J/\psi$ production at RHIC, leading to enhancement rather than suppression
in central collisions.  There is no evidence for a strong 
regeneration effect in the PHENIX Au+Au data so far \cite{rosatiproc}.  

\begin{figure}
\resizebox{0.45\textwidth}{!}{%
  \includegraphics{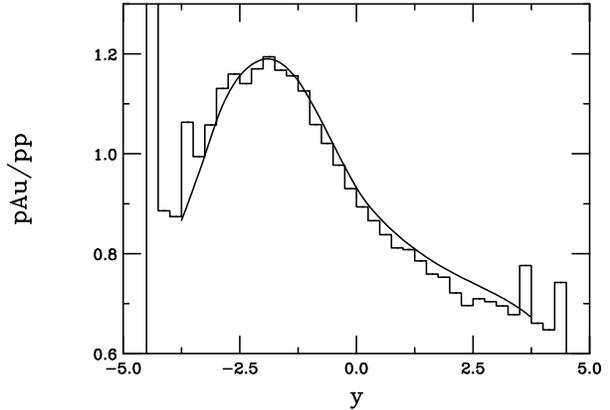}
}
\caption[]{ The $J/\psi$ $p$Au/$pp$ ratio at 200 GeV.  We compare the NLO 
(solid histogram, MRST HO) and LO (solid
curve, MRST LO) results using $m = \mu/2 = 1.2$ GeV
with the EKS98 parameterization.  
}
\label{psiratnlo}
\end{figure}

Shadowing, the modification of the parton densities in
the nucleus with respect to the free nucleon, may be taken into account by
replacing $f_j^p$ in Eq.~(\ref{sigpp}) by
$F_j^A(x,\mu^2,\vec b,z) = \rho_A(\vec b,z) S^j(A,x,\mu^2,\vec b,z) 
f_j^p(x,\mu^2)$.  The density distribution of the deuteron is also included
in these calculations but the small effects of shadowing in deuterium is 
ignored.  We did not discuss the effect of shadowing on
the charm $p_T$ distributions
because the effect at midrapidity is small and, on the logarithmic
scale of the $p_T$ distributions, negligible.  The $J/\psi$
is another story due to the PHENIX muon capability at forward and
backward rapidity.  As shown in Leitch's talk \cite{Leitchproc}, although
the PHENIX $J/\psi$ data are consistent with shadowing alone, the data
are also consistent with nuclear shadowing plus a small absorption
cross section of 1-3 mb, smaller than that currently obtained in SPS
measurements \cite{NA50talks}.  We have calculated $J/\psi$ production in
the color evaporation model (CEM) using the same mass and scale as in $c
\overline c$ production but cutting off the invariant mass of the pair at
$4m_D^2$.  The calculations of the d$A$/$pp$
ratios are done at LO to simplify the calculations.  As shown in 
Fig.~\ref{psiratnlo}, the LO and NLO ratios are equivalent.
We have now also implemented nucleon absorption in the calculation,
showing the effect of several absorption and production mechanisms.

To implement nuclear absorption on $J/\psi$ 
production in
$dA$ collisions, the $dN$ production cross section is weighted by the 
survival probability, $S^{\rm abs}$ \cite{rvherab} 
\begin{eqnarray}
S^{\rm abs}(\vec b,z) = \exp \left\{
-\int_{z}^{\infty} dz^{\prime}
\rho_A (\vec b,z^{\prime})
\sigma_{\rm abs}(z^{\prime} - z)\right\} \, \,
\label{nsurv}
\end{eqnarray}
where $z$ is the longitudinal production point and $z^{\prime}$ 
is the point at which the state is absorbed.
The nucleon absorption cross section, $\sigma_{\rm abs}$, typically
depends on where the
state is produced in the medium and how far it travels through nuclear matter.
If absorption alone is active, {\it i.e.}\ 
$S^j \equiv 1$, then an effective minimum bias $A$ dependence
is obtained after integrating $S^{\rm abs}$ over the spatial coordinates.
If $S^{\rm abs} = 1$ also, $\sigma_{{\rm d}A} = 2A \sigma_{pN}$.
When $S^{\rm abs} \neq 1$, $\sigma_{{\rm d}A} = 2A^\alpha \sigma_{pN}$ where,
if $\sigma_{\rm abs}$ is a constant, independent of the production mechanism
for a nucleus of $\rho_A = \rho_0 \theta(R_A - b)$,
$\alpha = 1 - 9\sigma_{\rm abs}/(16 \pi r_0^2)$,
where $r_0 = 1.2$ fm.  The contribution to the full $A$ 
dependence in $\alpha(x_F)$ from absorption alone is only constant if
$\sigma_{\rm abs}$ is constant and independent of the production mechanism
\cite{rvherab}.
The observed $J/\psi$ yield includes feed down from $\chi_{cJ}$ and $\psi'$
decays, giving
\begin{eqnarray} 
S_{J/\psi}^{\rm abs}(b,z) & = & 0.58 S_{J/\psi, \, {\rm dir}}^{\rm abs}(b,z) 
\nonumber \\
& + & 0.3 S_{\chi_{cJ}}^{\rm abs}(b,z) + 0.12 S_{\psi'}^{\rm abs}(b,z) 
\, \, . \label{psisurv} 
\end{eqnarray}  

\begin{figure*}
\resizebox{0.95\textwidth}{!}{%
  \includegraphics{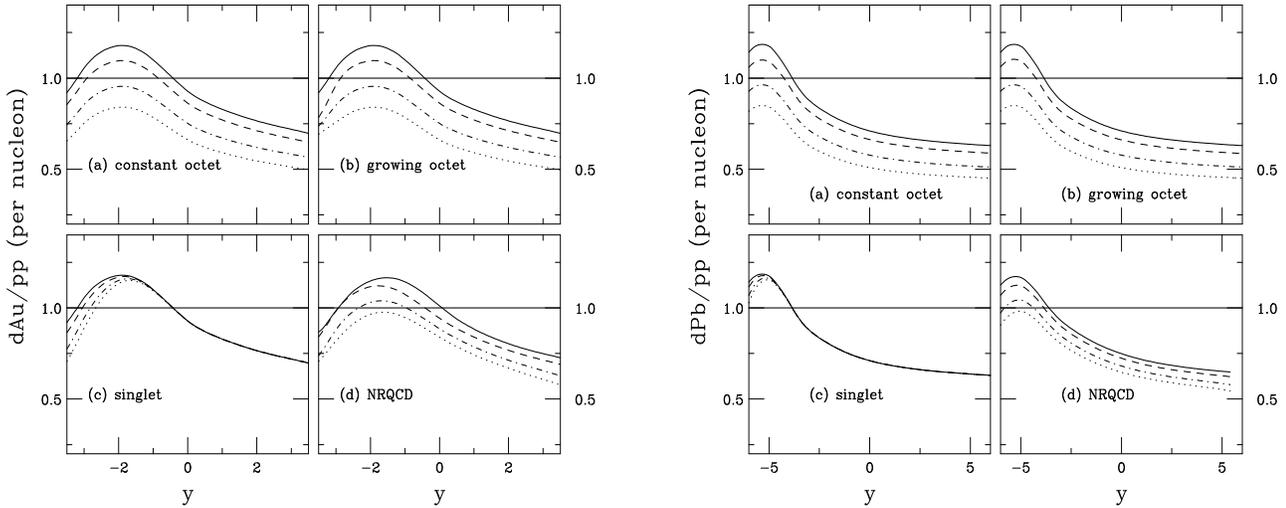}
}
\caption[]{The $J/\psi$ d$A$/$pp$ ratios with EKS98 for $\sqrt{S_{NN}} = 
200$ GeV (left-hand side) and for $\sqrt{S_{NN}} = 6.2$ TeV (right-hand side)
as a function of rapidity for (a) constant octet, (b) growing
octet, (c)
singlet, all calculated in the CEM and (d) NRQCD.  For (a)-(c),
the curves are no
absorption (solid), $\sigma_{\rm abs} = 1$ (dashed), 3 (dot-dashed)
and 5 mb (dotted).  For (d), we show no absorption
(solid), 1 mb octet/1 mb singlet
(dashed), 3 mb octet/3 mb singlet (dot-dashed), and 5 mb octet/3 mb singlet
(dotted).  
}
\label{abs}
\end{figure*}

The $J/\psi$ may be produced as a color singlet, a color octet or in a 
combination of the two.
In color singlet production, the final state absorption cross section 
depends on the size of the
$c \overline c$ pair as it traverses the nucleus, allowing absorption to be
effective only while the cross section is growing toward its asymptotic size
inside the target.
On the other hand, if the $c \overline c$ is only produced as a color octet, 
hadronization will occur only after the pair has traversed the target
except at very backward rapidity.  We have considered a constant octet cross
section, as well as one that reverts to a color singlet at backward rapidities.
For singlets, $S_{J/\psi, \, {\rm dir}}^{\rm abs} \neq 
S_{\chi_{cJ}}^{\rm abs} \neq S_{\psi'}^{\rm abs}$ but, with octets,
one assumes that $S_{J/\psi, \, {\rm dir}}^{\rm abs} = S_{\chi_{cJ}}^{\rm abs}
= S_{\psi'}^{\rm abs}$.
As can be seen in Fig.~\ref{abs}, 
the difference between the constant and growing octet assumptions
is quite small at large $\sqrt{S_{NN}}$ with only a small singlet effect 
at $y< -2$ and $-5$ at RHIC and the LHC
respectively.  Singlet absorption is also important only at similar
rapidities and is otherwise not different from shadowing alone.  
Finally, we have also considered a combination of octet and 
singlet absorption in the context of the NRQCD model, see Ref.~\cite{rvherab}
for more details.  The combination of nonperturbative singlet and octet 
parameters changes the shape of the shadowing ratio slightly.  Including the
singlet contribution weakens the effective absorption.  The results are
shown integrated over impact parameter.  The calculations use the EKS98 
shadowing parameterization
since it gives good agreement with the trend of the PHENIX data.  For results
with other shadowing parameterizations, see Ref.~\cite{rvphenix}.

Several values of the asymptotic
absorption cross section, $\sigma_{\rm abs} = 1$, 3 and
5 mb, corresponding to $\alpha = 0.98$, 0.95 and 0.92 respectively using
Eqs.~(\ref{nsurv}) and (\ref{psisurv})
are shown in
Fig.~\ref{abs}.  These values of
$\sigma_{\rm abs}$ are somewhat smaller than those obtained for the sharp
sphere approximation.
The diffuse surface of a real nucleus and the longer range of the
density distribution result in a smaller value of $\sigma_{\rm abs}$ than
a sharp sphere nucleus.  There is
good agreement with the trend of the preliminary PHENIX data \cite{phenixqm04}
for $\sigma_{\rm abs} = 0-3$ mb.
We use a value of 3 mb in our further
calculations to illustrate the relative importance of absorption and shadowing.

We now turn to the centrality dependence of $J/\psi$ production in d$A$ 
collisions.  In central collisions, inhomogeneous shadowing is stronger than 
the homogeneous result.  The
stronger the homogeneous shadowing, the larger the inhomogeneity.
In peripheral collisions, inhomogeneous
effects are weaker than the homogeneous results but some
shadowing is still present.  Shadowing persists in part because the
density in a heavy nucleus is large and approximately constant except
close to the surface and because the deuteron wave function has
a long tail.  We also expect absorption to be stronger in central
collisions.  
 
\begin{figure*}
\resizebox{0.95\textwidth}{!}{%
  \includegraphics{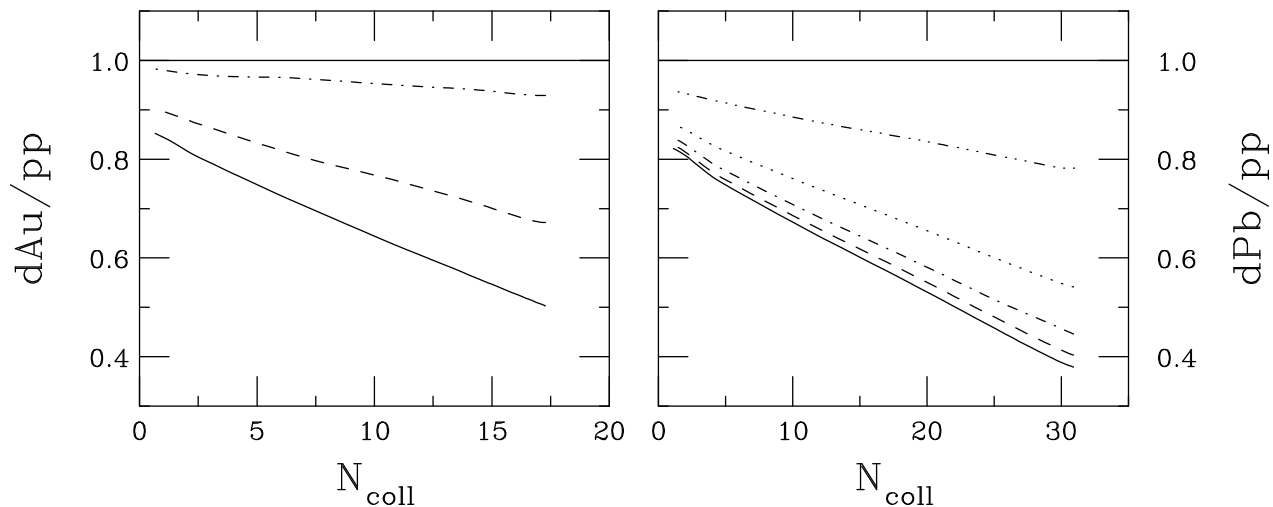}
}
\caption[]{The d$A$/$pp$ ratio as a function of $N_{\rm coll}$ for a growing
octet with $\sigma_{\rm abs} = 3$ mb and the EKS98 parameterization.  Left-hand
side: results for $y=-2$ (dot-dashed), $y=0$ (dashed)
and $y=2$ (solid) at 200 GeV.  Right-hand side:  results for $y = -4$ 
(dot-dot-dot-dashed), $y=-2$ (dotted), $y=0$ (dot-dashed), $y = 2$ (dashed)
and $y=4$ (solid) at 6.2 TeV. 
}
\label{ncoll_dep}
\end{figure*}

To study the centrality dependence of shadowing and absorption, we present the
d$A/pp$ ratios 
as a function of the number of binary $NN$ collisions, $N_{\rm coll}$,
\begin{eqnarray}
N_{\rm coll}(b) = \sigma_{NN}^{\rm in} \int d^2s T_A(s)
T_B (|\vec b - \vec s|) \nonumber
\end{eqnarray}
where $T_A$ and $T_B$ are the nuclear thickness functions and $\sigma_{NN}^{\rm
in}$ is the inelastic nucleon-nucleon cross section, 42 mb at 200 GeV and 76 mb
at 6.2 TeV.  In Fig.~\ref{ncoll_dep}, we show the $N_{\rm coll}$
dependence for several representative rapidities, $y = -2$, 0 and 2 for RHIC
(left-hand side) and $y= -4$, $-2$, 0, 2 and 4 for the LHC (right-hand side).
We have chosen an inhomogeneous shadowing parameterization proportional to the 
path length of the parton through the nucleus \cite{psidaprl}.  For results 
with other parameterizations, see Ref.~\cite{rvphenix}.

The dependence of the RHIC ratios on $N_{\rm coll}$ is almost linear, as seen
on the left-hand side of Fig.~\ref{ncoll_dep}.  
We do not show results for $N_{\rm coll} < 1$.  The weakest
$N_{\rm coll}$ dependence occurs in the antishadowing region, illustrated by
the $y = -2$ result (dot-dashed curve).  The trends of
the ratios as a function of $N_{\rm coll}$ are consistent with the PHENIX data
from the north muon arm ($y = 2$) and the electron arms ($y=0$)
but the preliminary
PHENIX results from the south arm ($y=-2$) are much stronger than
our predictions and, in fact, go the opposite way.
The overall dependence on $N_{\rm coll}$ is stronger than that
obtained from shadowing alone, described in Ref.~\cite{psidaprl} where
inhomogeneous shadowing effects depend
strongly on the amount of homogeneous shadowing.  Relatively large effects
at low $x$ are accompanied by the strongest $b$ dependence.  In the
transition region around midrapidity at RHIC, the $b$ dependence of
the ratio dAu/$pp$ due to shadowing is nearly
negligible and almost all the $N_{\rm coll}$ dependence at $y \sim 0$
can be attributed to absorption.  
The $y=-2$ results for color singlet production and absorption,
in the antishadowing
region, are fairly independent of $N_{\rm coll}$.
 
On the right-hand side of Fig.~\ref{ncoll_dep} we present our inhomogeneous
shadowing and absorption calculations for d+Pb collisions at $\sqrt{S_{NN}} =
6.2$ TeV at the LHC.  Results for $y = \pm 4$,
in the range of the ALICE muon arm, are also included.  
Given that the rapidity range of
the muon arm encompasses the crossover point where dPb/$pp \sim 1$ at $y \sim
-3.9$,
the centrality dependence of absorption alone could be determined and used to
calibrate the inhomogeneous shadowing effects.  Note that is is only
possible to reach $y \sim -3.9$ in ALICE by
running Pb+d since the muon arm is only on one side of midrapidity.
Both ALICE and CMS should be able to
measure $J/\psi$ production at $y = \pm 2$ and 0.
The results for $y = 4$, $\pm 2$ and 0, all in the low $x$ shadowing region,
are rather closely grouped together.  This should not be surprising because
the EKS98 shadowing ratios shown on the right-hand side of 
Fig.~\ref{abs} are not very strong
functions of rapidity.  Note that the $x$-axis scale is expanded
relative to that of RHIC due to the larger $\sigma_{NN}^{\rm in}$ at 
6.2 TeV.

\section{Summary}

In conclusion, the RHIC d+Au data on open charm and $J/\psi$ are beginning to
come into their own.  While the QCD calculations agree well with the shape of
the STAR $p_T$ distributions, they underestimate the reported total
cross section.  In contrast, the $J/\psi$ cross section is in relatively
good agreement with QCD predictions and the agreement of the
minimum bias data with calculations including shadowing and nucleon absorption 
is quite good.  The agreement of the $J/\psi$ calculations with the preliminary
PHENIX data is generally quite good, except for the dependence of the $y 
\approx -2$ results on $N_{\rm coll}$.

\end{document}